\documentclass[11pt]{article}
\usepackage{graphicx,amsmath,amssymb,fullpage,multirow}

\usepackage{multibbl}
\newbibliography{mt}
\newbibliography{ap}

\newcommand{\be}{\begin{equation}}
\newcommand{\ee}{\end{equation}}

\begin{document}

\title{Random Numbers Certified by Bell's Theorem}

\author{S. Pironio$^{1,2}\footnote{These authors contributed equally to this work.}$, A. Ac\'{\i}n$^{3,4*}$, S. Massar$^{1*}$, A. Boyer de la Giroday$^{5}$, D. N. Matsukevich$^{6}$,\\ P. Maunz$^{6}$, S. Olmschenk$^{6}$, D. Hayes$^{6}$, L. Luo$^{6}$, T. A. Manning$^{6}$, and C. Monroe$^{6}$\\[0.5em]
{\small$^1$Laboratoire d'Information Quantique, Universit\'e Libre de Bruxelles, 1050 Bruxelles, Belgium}\\
{\small$^2$Group of Applied Physics, University of Geneva, Geneva, Switzerland}\\
{\small$^3$ICFO-Institut de Ciencies Fotoniques, Castelldefels (Barcelona), Spain}\\
{\small$^4$ICREA-Institucio Catalana de Recerca i Estudis Avan\c{c}ats, 08010 Barcelona, Spain}\\
{\small$^5$Cavendish Laboratory, Cambridge CB3 0HE, United Kingdom}\\
{\small$^6$Joint Quantum Institute, University of Maryland Department of Physics and National}\\
{\small Institute of Standards and Technology, College Park, MD, 20742 U.S.A.}}
\date{}
\maketitle

Randomness is a fundamental feature in nature and a valuable
resource for applications ranging from cryptography and gambling
to numerical simulation of physical and biological systems. Random
numbers, however, are difficult to characterize
mathematically \cite{mt}{mrand}, and their generation must rely on an
unpredictable physical
process \cite{mt}{jennewein,stefanov,dynes,atsushi,fiorentino}.
Inaccuracies in the theoretical modelling of such processes or
failures of the devices, possibly due to adversarial attacks,
limit the reliability of random number generators in ways that are
difficult to control and detect.
Here, inspired by earlier work on nonlocality based \cite{mt}{ekert,barrett,masanes} and device independent \cite{mt}{mayers,magniez,acin,pironio,colbeck} quantum information processing, we show that the nonlocal correlations of entangled quantum particles can be used to certify the presence of genuine randomness. It is thereby possible to design of a new type of cryptographically secure random number generator which does not require any assumption on the internal working of the devices. This strong form of randomness generation is impossible classically and possible in quantum systems only if certified by a Bell inequality violation \cite{mt}{bell}.
We carry out a proof-of-concept demonstration of this
proposal in a system of two entangled atoms separated by
approximately 1 meter. The observed Bell inequality violation,
featuring near-perfect detection efficiency, guarantees that 42
new random numbers are generated with 99$\%$ confidence. Our
results lay the groundwork for future device-independent quantum
information experiments and for addressing fundamental issues
raised by the intrinsic randomness of quantum theory.
\newpage

The characterization of true randomness is elusive. There exist
statistical tests used to verify the absence of certain patterns
in a stream of numbers \cite{mt}{marsaglia,rukhin}, but no finite set
of tests can ever be considered complete, as there may be patterns
not covered by such tests. For example, certain pseudo-random
number generators are deterministic in nature, yet produce results
that satisfy all the randomness tests \cite{mt}{matsumoto}. At a more
fundamental level, there is no such thing as true randomness in
the classical world: any classical system admits in principle a
deterministic description and thus appears random to us as a
consequence of a lack of knowledge about its fundamental
description. Quantum theory is, on the other hand, fundamentally
random; yet, in any real experiment the intrinsic randomness of
quantum systems is necessarily mixed-up with an apparent
randomness that results from noise or lack of control of the
experiment. It is therefore unclear how to certify or quantify
unequivocally the observed random behaviour even of a quantum
process.

These considerations are of direct relevance to applications of randomness, and in particular cryptographic applications. Imperfections in random number generators \cite{mt}{jennewein,stefanov,dynes,atsushi,fiorentino} (RNG) can introduce patterns undetected by statistical tests but known to an adversary. Furthermore, if the device is not trusted but viewed as a black box prepared by an adversary, no existing RNGs can establish the presence of private randomness. Indeed, one can never exclude that the numbers were generated in advance by the adversary and copied into a memory located inside the device.

Here, we establish a fundamental link between the violation of
Bell inequalities and the unpredictable character of the outcomes
of quantum measurements and show, as originally proposed by
Colbeck \cite{mt}{colbeck}, that the nonlocal correlations of quantum
states can be used to generate certified private randomness. The
violation of a Bell inequality \cite{mt}{bell} guarantees that the
observed outputs are not predetermined and that they arise from
entangled quantum systems that possess intrinsic randomness.
For simplicity, we consider the Clauser-Horn-Shimony-Holt (CHSH)
form of Bell inequality \cite{mt}{clauser}, but our approach is general
and applies to any Bell inequality. We thus consider two separate
systems that can each be measured in two different ways, with a
measurement on each system resulting in one of two values (Figure
1). The binary variables $x$ and $y$ describe the type of
measurement performed on each system, resulting in respective
binary measurement outcomes $a$ and $b$. We quantify the Bell
inequality violation through the CHSH correlation function \cite{mt}{clauser}
\be\label{chsh}%\hat I = \sum_{x,y} (-1)^{xy} \frac{N(a=b,xy)-N(a\neq b,xy)}{nP(xy)}\,,
I=\sum_{x,y}(-1)^{xy}\left[P(a=b|xy)-P(a\neq b|xy)\right]\,,
\ee
where $P(a=b | xy)$ is the probability that $a=b$
given settings $(x,y)$ and $P(a \ne b | xy)$ is defined
analogously.
Systems that admit a local, hence deterministic \cite{mt}{fine},
description satisfy $I\leq 2$. Certain measurements performed on
entangled states, however, can violate this inequality.

In order to estimate the Bell violation, the experiment is
repeated $n$ times in succession. The measurement choices $(x,y)$
for each trial are generated by an identical and independent
probability distribution $P(xy)$. We denote the final output
string after the $n$ runs $r=(a_1,b_1;\ldots; a_n,b_n)$ and the
input string $s=(x_1,y_1;\ldots; x_n,y_n)$. An estimator $\hat I$ of the
CHSH expression\eqref{chsh} determined
from the observed data is
given by
\be\label{chsh2} \hat I = \frac{1}{n}\sum_{x,y} (-1)^{xy} \left[N(a=b,xy)-N(a\neq b,xy)\right]/P(xy)\,, \ee
%$\hat I = \sum_{x,y} (-1)^{xy} \left[N(a=b,xy)-N(a\neq b,xy)\right]/[nP(xy)]$
where
$N(a=b,xy)$ is the number of times that the measurements $x,y$
were performed and that the outcomes $a$ and $b$ were found equal
after $n$ realizations, and where $N(a\neq b,xy)$ is defined
analogously.

The randomness of the output string $r$ can be quantified by the
min-entropy \cite{mt}{renner}
$H_\infty(R|S)= -\log_2\left[\max_rP(r|s)\right]$, where $P(r|s)$ is the conditional
probability of obtaining the outcomes $r$ when the measurements
$s$ are made. We show (see Appendix A) that the
min-entropy of the outputs $r$ is bounded by \be H_\infty(R|S)\geq
nf(\hat I-\epsilon)\label{bound2} \ee with probability greater
than $1-\delta$, where $\epsilon=O(\sqrt{-\log\delta/(q^2n)})$ is
a statistical parameter  and $q=\min_{x,y} P(xy)$ is the
probability of the least probable input pair. The function $f(I)$
is obtained using semidefinite
programming \cite{mt}{navascues1,navascues2} and presented in Figure~2.
To derive the bound~(\ref{bound2}), we make the following
assumptions: $1)$ the two observed systems satisfy the laws of
quantum theory; $2)$ they are separated and non-interacting during
each measurement step $i$; $3)$ the inputs $x_i,y_i$ are generated
by random processes that are independent and uncorrelated from the
systems and their value is revealed to the systems only at step
$i$ (see Figure~1). Other than these assumptions, no constraints
are imposed on the states, measurements, or the dimension of the
Hilbert space. We do not even assume that the system behaves
identically and independently for each trial; for instance, the
devices may have an internal memory (possibly quantum), so that
the $i^\text{th}$ measurement can depend on the previous $i-1$
results and measurement settings. Any value of the min-entropy
smaller than (\ref{bound2}) is incompatible with quantum theory.
The observed CHSH quantity $\hat I$ thus provides a bound (see
Figure~3) on the randomness produced by the quantum devices,
independent of any apparent randomness that could arise from noise
or limited control over the experiment.

This result can be exploited to construct a novel RNG where the
violation of a Bell inequality guarantees that the output is
random and private from any adversary, even in a
device-independent scenario \cite{mt}{acin,pironio} where the internal
workings of the two quantum devices are unknown or not trusted
(See Appendix~B). Some amount of randomness at
the inputs is necessary to perform the statistical tests used to
estimate the Bell inequality violation. Hence what we describe
here is a randomness expansion scheme \cite{mt}{colbeck}
where a small private random seed is expanded into a longer
private random string.
%(The situation is analoguous to Quantum Key Distribution which should more properly be called Quantum Key Expansion, since some initial key is necessary to authenticate the public communication channel.)
The randomness used to choose the inputs needs not be divulged and
can be used subsequently for another task. The final random
string, the concatenation of the input and output random strings,
is thus manifestly longer than the initial one. However, when $n$
becomes sufficiently large, a catalysis effect is possible wherein
a seed string of length $O(\sqrt{n}\log{\sqrt{n}})$ produces a
much longer random output string of entropy $O(n)$, as illustrated
in Figure 3 (see Appendix B). This is possible
because $I$ can be adequately estimated without consuming much
randomness by using the same input pair most of the time [e.g.,
$(x,y)=(0,0)$] while only seldom sampling from the other
possibilities, in which case $q\ll 1$. This follows from the fact
that the CHSH expression depends only the conditional
probabilities $P(ab|xy)$, which can be estimated even if $x,y$ are
not uniformly distributed.

Although the final output string may not be uniformly random (it may not even pass one of the usual statistical tests \cite{mt}{marsaglia,rukhin} of randomness), we are guaranteed that its entropy is bounded by (\ref{bound2}). With the help of a small private random seed,  the output string can then be classically processed using a randomness extractor \cite{mt}{nisan} to convert it into a string of size $nf(\hat I-\epsilon)$ that is nearly uniform and uncorrelated to the information of an adversary.
The bound~(\ref{bound2}) establishes security of our randomness expansion protocol against an adversary that measures his side-information before the randomness extraction step, e.g., against an adversary that has only a short-lived or bounded quantum memory.
This is because it applies when conditioned to any measurement performed by the adversary.
However, our protocol is not yet proven to be universally-composable against a full quantum adversary, that is, secure against an adversary that stores his side-information in a quantum memory which can be measured at a later stage. A universally-composable proof would also cover the situation in which the adversary tries to estimate the random numbers after getting partial information about them. Proving universally-composable security of our protocol would also probably lead to much more efficient randomness expansion schemes. Note that the fact that the bound~(\ref{bound2}) holds for devices that have an internal memory is a significant advance over the device-independent protocols \cite{mt}{masanes,acin,pironio} proposed so far. It is the crucial feature that makes our protocol practical.

The experimental realization of this scheme requires the
observation of a Bell inequality with the detection loophole
closed (near-perfect detection of every event), so that the
outputs $r$ cannot be deterministically reproduced. The two
individual systems should also be sufficiently separated so that
they do not interact, but it is not necessary for the two
subsystems to be space-like separated (see Appendix~C).

We realize this situation with two $^{171}$Yb$^+$ atomic ion
quantum bits (qubits) \cite{mt}{Yb} confined in two independent vacuum
chambers separated by about 1 meter. The qubit levels within each
atom are entangled through a probabilistic process whereby each
atom is entangled with emitted photons and the interference and
coincident detection of the two photons heralds successful
preparation of a near-maximal entangled state of the two remote
atomic qubits through entanglement swapping \cite{mt}{matsukevich}, as
described in Fig. 1 and Appendix~D. The binary
values $a$ and $b$ correspond to subsequent measurement results of
each qubit obtained through standard atomic fluorescence
techniques (detection error $<3\%$) \cite{mt}{Yb}, and every event is
recorded. The respective binary measurement bases $x$ and $y$ are
chosen randomly and set by coherent qubit rotation operations
prior to measurement. Direct interaction between the atoms is
negligible and classical microwave and optical fields used to
perform rotations and measurements on one atom have no influence
on the other atom (we perform statistical tests to verify that
the measurement data is compatible with this hypothesis, see
Appendix~D.4). To estimate the value of the CHSH
inequality we accumulate $n = 3,016$ successive entanglement
events over the period of about one month, summarized in
Appendix~D.1 and Table I. The observed CHSH
violation of $\hat I=2.414$ represents a substantial improvement
over previous results \cite{mt}{rowe,matsukevich}. The probability that
a local theory, possibly including an internal memory of past
events \cite{mt}{gill2}, could produce such a violation is $P(\hat
I\geq 2.414)\leq 0.00077$ (see Appedendix~D.3).
\begin{table*}[tb]
\begin{tabular}{|c| c| c c c c|c c|}
\hline Inputs & Rotations &  \multirow{2}*{$N(0,0;x,y)$} &   \multirow{2}*{$N(0,1;x,y)$} & \multirow{2}*{$N(1,0;x,y)$} & \multirow{2}*{$N(1,1;x,y)$} & Total & \multirow{2}*{$P(a=b | xy)$}\\
$(x,y)$ & $(\varphi_x, \varphi_y)$ & & & & & events&\\
\hline
\centering{0,0} & \centering{$0^\circ$, $45^\circ$}     &293&    94&    70&    295& \centering{752}&    0.782\\
\centering{0,1} &\centering{$0^\circ$, $135^\circ$}     &298&    70&    74 &309&    \centering{751}&    0.808\\
\centering{1,0}&    \centering{$90^\circ$, $45^\circ$}&     283 &69 &   64&    291& \centering{707}&    0.812\\
\centering{1,1} &\centering{$90^\circ$, $135^\circ$} &  68&    340 &  309  &89& \centering{806}&
0.195\\\hline
\end{tabular}
\caption{Observed number of events $N(a,b;x,y)$ for which  the measurement on one atom gave outcome $a$ and the measurement on the other atom gave outcome $b$, given the binary choices of the measurement bases $(x,y)$ corresponding to $\pi/2$ qubit rotations with phase angles $(\varphi_x,\varphi_y)$ on the equator of the Bloch sphere.  The last column gives the fraction of events where $a=b$ given each input.  If the experiment is interpreted as consisting of identical and independent realisations (an assumption not made elsewhere in this paper), the data then indicate a CHSH observable of
$\hat{I}=\sum_{x,y}(-1)^{xy}[P(a = b|xy)-P(a \neq b|xy)]=2.414\pm0.058$,
significantly beyond the local-deterministic threshold of $I = 2$.}
\end{table*}
In the experiment, we chose a uniform random distribution of the
initial measurement bases inputs [$P(x,y) = 1/4$] to minimize the
number of runs required to obtain a meaningful bound on the output
entropy (see Figure 3). The observed CHSH violation implies that
at least $H(R|S) > 42$ new random bits are generated in the
experiment with a $99\%$ confidence level.
This is the first time that one can certify that new randomness is produced in an experiment without a detailed model of the devices. We rely only on a high-level description (atoms confined to independent vacuum chambers separated by one meter) to ensure the absence of interaction between the two subsystems when the measurements are performed.  Since no active measures are taken in our experiment to control this interaction, these new random bits cannot be considered private in the strongest adversarial device-independent scenario. The level of security provided by our experiment will nevertheless be sufficient for many applications as it guarantees that almost all failure modes of the devices will be detected.
The current experiment does not reach the catalysis regime
mentioned above, owing to the low success probability of heralded
entanglement generation ($2 \times 10^{-8}$) [26].
However, it should be possible to exceed the catalysis threshold
by improving the photon-collection efficiency through the use of
nearby optical collection elements or optical cavities \cite{mt}{luo}.

Stepping back to the more conceptual level, note that
Eq.~(\ref{bound2}) relates the random character of quantum theory
to the violation of Bell inequalities. This bound can be modified
for a situation where we assume only the no-signalling principle
instead of the entire quantum formalism (see Figure 2 and 3 and
Appedenix~A.3). Such a bound lays the basis for
addressing in a statistically significant way one of the most
fundamental questions raised by quantum theory: whether our world
is compatible with determinism (but then necessarily signalling),
or inherently random (if signalling is deemed impossible).

{\textbf{Acknowledgements} We thank Roger Colbeck for sharing his PhD thesis with us.
This work was supported by the Swiss NCCR \textit{Quantum Photonics}, the European ERC-AG \textit{QORE}, the European QAP, COMPAS projects and ERC Starting grant PERCENT, the Spanish MEC FIS2007-60182 and Consolider-Ingenio QOIT projects, Generalitat de Catalunya and Caixa Manresa, the Interuniversity Attraction Poles Photonics@be Programme (Belgian Science Policy), the U.S. Army Research Office with funds from IARPA, the National Science Foundation (NSF) Physics at the Information Frontier Program, and the NSF Physics Frontier Center at JQI.}

\begin{figure}
    \centering
\includegraphics[width=1\columnwidth,keepaspectratio]{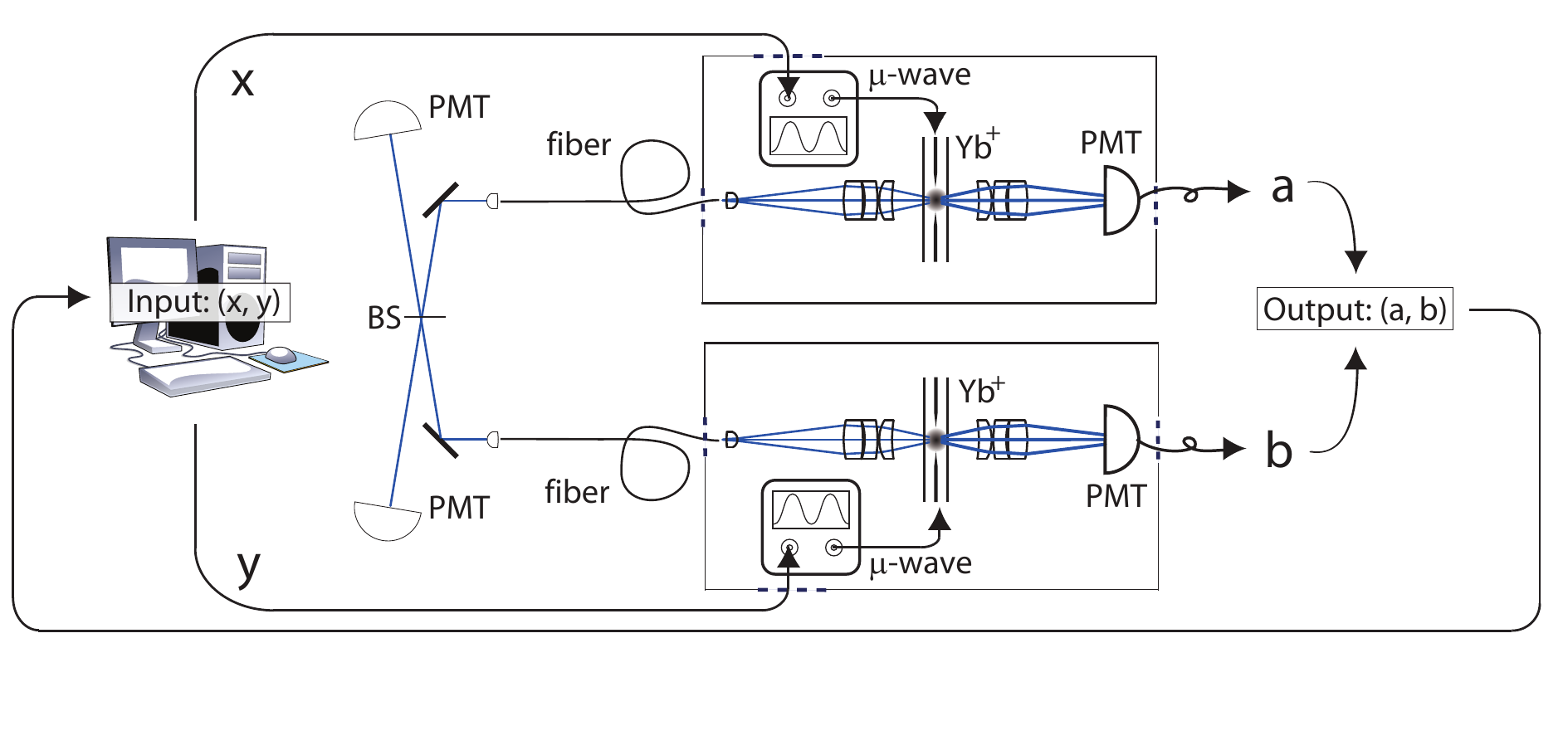}
    \caption{Experimental realization of private random number generator using two $^{171}$Yb$^+$ qubits trapped in independent vacuum chambers. Each atom emits a single photon (to the left) that is entangled with its host atomic qubit and coupled into optical fibers; the interference of the photons on the beamsplitter (BS) and subsequent coincidence detection on photomultiplier tubes (PMT) herald the entanglement of the atomic qubits \cite{mt}{matsukevich}.  After the qubits are entangled, binary random inputs $(x,y)$ are fed to microwave ($\mu$-wave) oscillators that coherently rotate each qubit in one of two ways prior to measurement \cite{mt}{matsukevich}.  Each qubit is finally measured through fluorescence that is collected on PMTs \cite{mt}{Yb}, resulting in the binary outputs $(a,b)$. Abstractly, we can view this scheme as composed of two black boxes that receive inputs $x,y$ and produce outputs $a,b$. In our theoretical analysis, no hypothesis are made about the internal working of the devices, but the classical and quantum information flowing in and out of the boxes is restricted (dotted lines). In particular, the two boxes are free to communicate before inputs are introduced (to establish shared entanglement), but are no longer allowed to interact during the measurement process. Moreover the values of the inputs are revealed to the boxes only at the beginning of the measurement process. In the experiment,
no active measures are taken to control the flow of information in and out of the systems. However, once the atoms are entangled, direct interaction between them is negligible. Moreover the value of the chosen measurement bases $(x,y)$, obtained by combining the outputs of several random number generators, is unlikely to be correlated to the state of the atoms before the measurement microwave pulses are applied, satisfying the conditions for the bound (\ref{bound2}) on the entropy of the outputs.}
    \label{fig2}
\end{figure}
\clearpage\begin{figure}
    \centering
  \includegraphics[width=0.8\columnwidth,keepaspectratio]{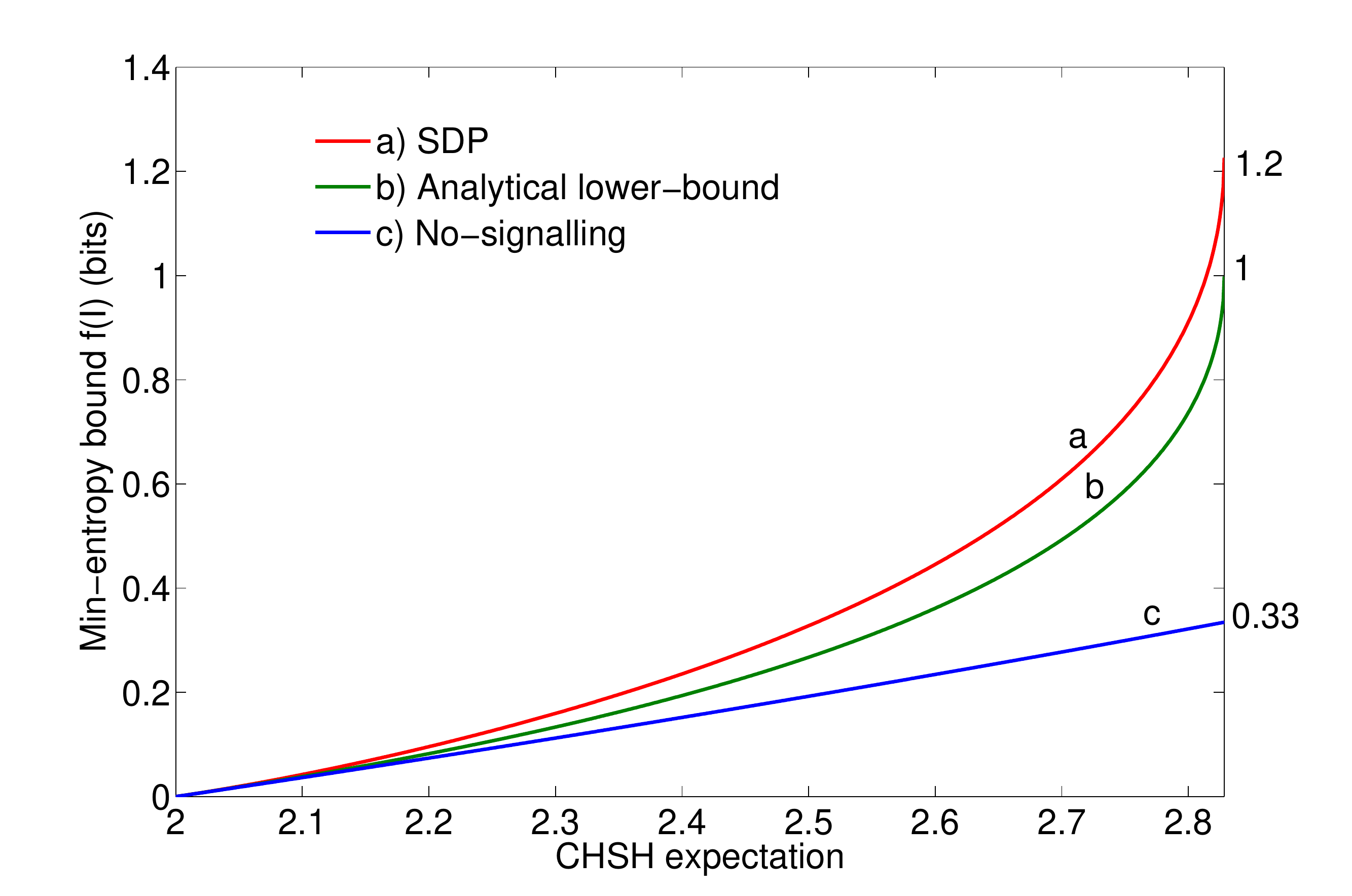}
    \caption{Plot of the function $f(I)$. The function $f(I)$ can be interpreted as a bound on the min-entropy per use of the system for a given CHSH expectation $I$, in the asymptotic limit of large $n$ where finite statistics effects (the parameter $\epsilon$ in (\ref{bound2})) can be neglected. The function $f(I)$ (curve $a$) is derived through semidefinite programming using the techniques of [22,23] (semidefinite programming is a numerical method that is guaranteed to converge to the exact result). Curve $b$ corresponds to the analytical lower-bound $f(I)\geq -\log_2\left[1-\log _{2} \left(1+\sqrt{2-\frac{I^{2} }{4} } \right)\right]$.  Curve $c$ corresponds to the minimal value $f(I)=-\log_2\left(3/2-I/4\right)$ of the min-entropy implied by the no-signalling principle alone. The function $f(I)$ starts at zero at the local threshold value $I=2$. Systems that violate the CHSH inequality ($I>2$), on the other hand, satisfy $f(I)>0$, i.e., have a positive min-entropy.}
    \label{fig3}
\end{figure}
\begin{figure}
  \centering
  \includegraphics[width=0.8\columnwidth,keepaspectratio]{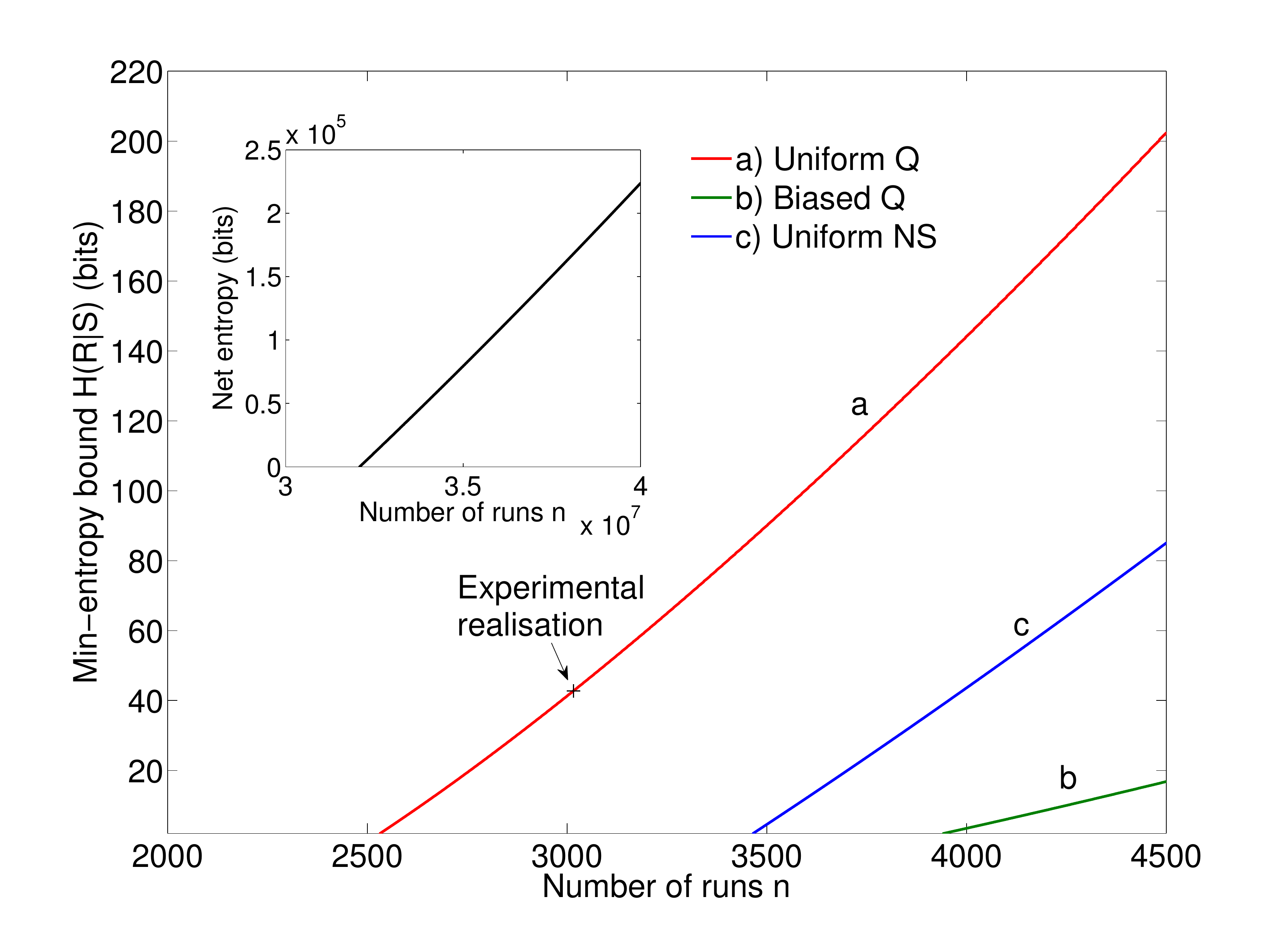}
    \caption{Bound $nf(I)$ on the minimum entropy produced vs. the number of trials $n$ for an observed CHSH violation of $\hat I=2.414$, and a confidence level $1-\delta = 99\%$.  The amount of randomness given by the bound~(\ref{bound2}) depends on the probability with which the inputs of each trial $(x_i,y_i)$ are chosen through the parameter $q=\min_{x,y}[P(x,y)]$, where $P(x,y)$ is the probability distribution of the inputs. We have plotted the bounds on the entropy implied by quantum theory for a uniform choice of inputs [$P(x,y) = 1/4$] (curve $a$) and for a biased choice of inputs given by $P(00) = 1-3q, P(01) = P(10) = P(11) = q$, where $q = \alpha n^{-1/2}$ with $\alpha=11$ (curve $b$). For a given number $n$ of uses of the devices, the uniform choice of inputs leads to more randomness in the outputs. On the other hand, biased inputs require less randomness to be generated, and the net amount of randomness produced given by the difference between the output and input entropy becomes positive for sufficiently large $n$. Curve $c$ represents the bound on the entropy implied by the no-signalling principle alone for a uniform choice of inputs.  Note that in all cases, a minimal number of uses of the devices (a few thousand) is required to guarantee that some randomness has been produced at the confidence level $1-\delta=99\%$ The inset shows the net amount of entropy produced (output entropy minus input entropy) for the biased choice of inputs with the observed CHSH violation.}
    \label{fig4}
\end{figure}

\clearpage\bibliographystyle{mt}{unsrt}
\bibliography{mt}{randomness_mt}{References}

\newpage
\appendix
\part*{Appendix}
\tableofcontents
\section{Theoretical results}
In this section, we derive the bound~(3) presented in the main text. We follow a two-step approach. In Section~A.1, we quantify the randomness of a pair of quantum systems characterized by a given Bell expectation $I$. In Section~A.2, we apply these results to the experimental data produced by Bell-violating devices that are used $n$ times in succession and which, in full generality, may not behave identically and independently at each trial.

\subsection{Quantum randomness versus Bell violation}\label{qvb}
We establish here a relation between the randomness of the measurement outcomes of a quantum system and its expected Bell violation. In full generality, we consider a Bell scenario with $m$ different measurements per system, each measurement having $d$ possible results. The two quantum subsystems are characterized by the joint probabilities $P_{AB|XY} =\left\{P(ab|xy)\right\}$ to produce outcomes $a,b=1,\ldots ,d$, when measurements $x,y=1,\ldots ,m$ are made. A Bell expression associated to this scenario is a linear combination $I=\sum _{abxy}c_{abxy}  P(ab|xy)$ of the probabilities specified by the $m^{2} \times d^{2} $ coefficients $\{ c_{abxy} \} $. Probabilities that admit a local description satisfy $I\le I_{0} $, where $I_{0} $ is the local bound of the Bell inequality. The CHSH correlation function (1) is a particular example of such a Bell expression with $m=2$ and $d=2$.

We quantify the randomness of the output pairs conditioned on the input pairs by the min-entropy $H_\infty(AB|XY)=$$-\log _{2} \max _{ab} P(ab|xy)$$=\min _{ab} $ $\left[-\log _{2} P(ab|xy)\right]$.
For given Bell violation $I$, our aim is to obtain a lower bound on the min-entropy
\begin{equation} \label{eq1}
H_\infty(AB|XY)\ge f(I)
\end{equation}
satisfied by all quantum realizations of the Bell scenario. Let $P^{*} (ab|xy)$ denote the solution to the following optimization problem:
\begin{equation}
\label{eq2}
\begin{array}{rcc}
P^*(ab|xy)=&\max &P(ab|xy)\\
& \text{subject to}&\sum_{abxy}c_{abxy}P(ab|xy)=I\\
&&P(ab|xy)={\text{tr}}\left(\rho M_{x}^{a} \otimes M_{y}^{b} \right)
\end{array}
\end{equation}
where the optimization is carried over all states $\rho $ and all measurement operators $M_{x}^{a} $ and $M_{y}^{b} $, defined over Hilbert spaces of arbitrary dimension. The minimal value of the min-entropy compatible with the Bell violation $I$ and quantum theory is then given by $H_\infty(AB|XY)=-\log _{2} \max _{ab} P^{*} (ab|xy)$.

To obtain a lower-bound on the min-entropy as a function of the Bell violation $I$ which does not depend on the input pair $(x,y)$, it is thus sufficient to solve (\ref{eq2}) for all output and input pairs $(a,b)$ and $(x,y)$. This can be done by adapting in a straightforward way the technique introduced in \cite{ap}{apnavascues1,apnavascues2} and generalized in \cite{ap}{appironio2}.

In \cite{ap}{apnavascues1, apnavascues2}, an infinite hierarchy of conditions $C_{i} $ ($i=1,2,\ldots $) necessarily satisfied by all probabilities of the form $P(ab|xy)={\text{tr}}\left(\rho M_{x}^{a} \otimes M_{y}^{b} \right)$ are introduced. These conditions can be characterized by semidefinite programs (SDP) of finite size. Conditions higher in the hierarchy, that is corresponding to higher values of $i$, are more constraining, i.e., they better reflect the constraints $P(ab|xy)={\text{tr}}\left(\rho M_{x}^{a} \otimes M_{y}^{b} \right)$, but they also correspond to SDP's of larger size. By replacing in \eqref{eq2} the conditions $P(ab|xy)={\text{tr}}\left(\rho M_{x}^{a} \otimes M_{y}^{b} \right)$ by the semidefinite constraints $C_{i} $ for $i=1,2,\ldots $, one obtains a sequence of relaxations $R_{1} ,R_{2} ,\ldots $ of the original problem that can be solved through semidefinite programming since the constraint $\sum _{abxy}c_{abxy}  P(ab|xy)=I$ and the objective function in \eqref{eq2} are both linear in the probabilities $P(ab|xy)$. The solutions of these relaxations yield a sequence of upper-bounds on the optimum of \eqref{eq2}, which in turn yields a sequence of lower bounds on the min-entropy: $f^{(1)} (I)\le f^{(2)} (I)\le \ldots \le H_\infty(AB|XY)$. The exact form of these lower bounds depend on the Bell inequality specified by the coefficients $\{ c_{abxy} \} $, but they are all convex functions that are equal to 0 at the classical point $I=I_{0} $.

Formulating the problem as an SDP is extremely useful, as numerical methods are guaranteed to find the global optimum of the SDP, i.e., will yield up to numerical precision the exact values of $f^{(1)} (I),f^{(2)} (I),\ldots $.We have solved the SDPs corresponding to the second relaxation step ($i=2$) for the CHSH inequality using the matlab toolboxes SeDuMi~\cite{ap}{apsedumi} and YALMIP~\cite{ap}{apyalmip} (see Figure 2).

Upper bounds on the min-entropy can be obtained by searching numerically for solutions to \eqref{eq2} with fixed Hilbert space-dimension. Specifically, one can introduce a parameterization of the state and measurement operators and vary the parameters to maximize \eqref{eq2}. We find that our upper and lower bounds coincide up to numerical precision ($10^{-9} $), that is, the relation between the min-entropy and the CHSH violation presented in Figure 2 is tight.

Using the same method, it is also possible to find a relation between the Bell violation $I$ and the local min-entropy, defined by $H_\infty(A|X)=-\log _{2} \max _{a,x} P(a|x)$, where $P(a|x)=\sum _{b}P (ab|xy)$ is the marginal probability distribution of Alice's system. Note that the local min-entropy gives a lower bound on the global min-entropy, since $H_\infty(AB|XY)\ge H_\infty(A|X)$.

In the case of the CHSH inequality, we are able to obtain the following tight analytical lower bound (see Figure 2):
\begin{equation} \label{eq3}
H_\infty(A|X)\ge 1-\log _{2} \left(1+\sqrt{2-\frac{I^{2} }{4} } \right)\, .
\end{equation}
To derive the above bound we use the fact that the in the case of the CHSH inequality (two inputs with binary outcomes), it is sufficient to consider measurements on a two-qubit system~\cite{ap}{appironio} of the form $|\Psi _{\theta } \rangle =\cos \theta |00\rangle +\sin \theta |11\rangle $. For a given state $|\Psi _{\theta } \rangle $, one can then easily compute both $P_{\theta }^{*} =\max _{a,x} P_{\theta } (a|x)\le \cos 2\theta $, and using the Horodecki criterion~\cite{ap}{aphorodecki}, the maximal CHSH violation $I_{\theta } \le 2\sqrt{1+\sin ^{2} \theta } $. The region that is accessible to quantum theory, is then given by the convex hull of the above set of points when we let $0\le \theta \le \pi /4$. This corresponds to the region characterized by the inequality $P(a|x)\le \frac{1}{2} \left(1+\sqrt{2-I^{2} /4} \right)$, which implies Eq. (\ref{eq3}). The above bound is tight because it is possible to achieve simultaneously $P_{\theta }^{*} =\cos 2\theta $ and $I_{\theta } =2\sqrt{1+\sin ^{2} \theta } $ with the same set of measurements on the state $|\Psi _{\theta } \rangle $.

\subsection{Randomness produced by Bell devices used $n$ times in succession}\label{n}
To apply the results of the previous section to the experimental data produced by devices that violate a Bell inequality, one has first to estimate the Bell violation. This requires to use the devices a large number $n$ of times in succession. In full generality,  one cannot assume, however,  that the devices behave identically and independently at each use. For instance, they may have an internal memory, so that what happens at the $i^\text{th}$ use of the devices depends on what happened during the $i-1$ previous uses. The results derived in the previous section must therefore be combined with a statistical approach that takes into account such memory effects. We carry out this analysis below and show how the randomness produced by the devices can be quantified and determined experimentally without making any particular assumptions about their internal behaviour.

We thus suppose that we have devices that violate a Bell inequality. The devices are used $n$ times in succession. We denote by $x_{i} ,y_{i} \in \{ 1,\ldots ,m\} $ the measurement inputs and by $a_{i} ,b_{i} \in \{ 1,\ldots ,d\} $ the measurements outputs at round $i$. We denote by $a^{k} =(a_{1} ,a_{2} ,\ldots ,a_{k} )$ the string of the first $k$ outputs $a_{i} $, and we define similarly $b^{k} $, $x^{k} $, and $y^{k} $. We suppose that the input pairs ($x_{i} ,y_{i} $) at round $i$ are independent and identical random variables with distribution $P(x_{i}=x ,y_{i}=y )=P(x,y)$. Note that in general $P(x,y)$ may not be a product distribution $P(x,y)=P(x)P(y)$.

Let $P_{R|S} =\left\{P(a^{n} b^{n} |x^{n} y^{n} )\right\}$ be the probability distribution of the final output string $r=(a^{n} ,b^{n} )$ given the fact that the sequence of inputs $s=(x^{n} ,y^{n} )$ has been inserted in the devices. Similarly to the single-copy case, the randomness of the output string conditioned on the inputs can be characterized by the min-entropy $H_\infty(R|S)=$$-\log _{2} \max _{r} P(r|s)$$=\min _{a^{n} b^{n}} $ $\left(-\log _{2} P(a^{n} b^{n} |x^{n} y^{n} )\right)$. We derive here a lower bound on $-\log _{2} P(a^{n} b^{n} |x^{n} y^{n} )$ valid for all $(a^{n} b^{n} $) and $(x^{n} ,y^{n} )$. This implies a lower bound on $H_\infty(R|S)$, as $H_\infty(R|S)\ge -\log _{2} P(a^{n} b^{n} |x^{n} y^{n} )$.  We start with the identities
\begin{eqnarray} \label{eq4}
-\log_2 P(a^nb^n|x^ny^n)&=&-\log_2\prod_{i=1}^n P(a_ib_i|a^{i-1}b^{i-1}x^iy^i)\nonumber\\
&=&-\log_2\prod_{i=1}^n P(a_ib_i|x_iy_iW^i)\\
&=&\sum_{i=1}^n - \log_2 P(a_ib_i|x_iy_iW^i)\nonumber
\end{eqnarray}
The first equality follows from Bayes rule together with the fact that the response of the system at round $i$ does not depend on future inputs $(x_j,y_j)$ with $j>i$. In the second equality, we introduced the variable $W^{i} =(a^{i-1} b^{i-1} x^{i-1} y^{i-1} )$ to denote all events in the past of round $i$. The behaviour of the devices at round $i$ conditioned on the past is characterized by a response function $P(a_{i} b_{i} |x_{i} y_{i} W^{i} )$ and a Bell violation $I(W^{i} )$. Whatever be the precise form of the quantum state and measurements implementing this behaviour, they are bound to satisfy the constraint $-\log _{2} P(a_{i} b_{i} |x_{i} y_{i} W^{i} )\ge f(I(W^{i} ))$, as required by \eqref{eq1}. Inserting this relation into Eq. \eqref{eq4}, we obtain
\begin{eqnarray} \label{eq5}
-\log_2 P(a^nb^n|x^ny^n)&\geq&\sum_{i=1}^n f\left(I(W^i)\right)\nonumber\\
&\geq&nf\left(\frac{1}{n}\sum_{i=1}^nI(W^i)\right)
\end{eqnarray}
where we used the convexity of the function $f$ to deduce the second inequality. We now show that the quantity $\frac{1}{n} \sum _{i=1}^{n}I (W^{i} )$ can be estimated with precision using only the information coming from the input and output strings observed in the experiment.

Let $\chi (e)$ be the indicator function for the event $e$, i.e. $\chi (e)=1$ if the event $e$ is observed, $\chi (e)=0$ otherwise. Consider the random variable
\begin{equation} \label{eq6}
\hat{I}_{i} =\sum _{abxy}c_{abxy}  \frac{\chi (a_{i} =a,b_{i} =b,x_{i} =x,y_{i} =y)}{P(x,y)} ,
\end{equation}
which is chosen so that its expectation conditioned on the past $W^{i} $ is equal to $E(\hat{I}_{i} |W^{i} )=I(W^{i} )$. We define $\hat{I}=\frac{1}{n} \sum _{i=1}^{n}\hat{I}_{i}  $ as our estimator of the Bell violation. In the case of the CHSH inequality, it is readily verified that (\ref{eq6}) correspond to the expression (2) given in the main text.
Let $q=\min _{xy} \{P(x,y)\} $ and let us assume that $q>0$ (that is, we assume that each possible pair of inputs $(x,y)$ has a non-zero probability to be chosen).  Let us introduce the random variables $Z^{k} =\sum _{i=1}^{k}\left(I_{i} -I(W^{i} )\right) $.  It can be verified that $(i)$ $E(|Z^{k} |)<\infty $, and that $(ii)$ $E(Z^{k} |W^{1} ,\ldots ,W^{j} )=E(Z^{k} |W^{j} )=Z^{j} $ for $j\le k$. Thus the sequence $\{ Z^{k} \, :\, k\ge 1\} $ is a martingale\footnote{
A martingale is a stochastic process (i.e., a sequence of random variables) such that the conditional expected value of an observation at some time $k$, given all the observations up to some earlier time $j$, is equal to the observation at that earlier time $j$. This causality constraint --- expressed formally by condition $(ii)$ above --- is very strong, and it implies that in some respects martingales behave like sums of independent variables. In particular, if the martingale increments (from time $k$ to the time $k+1$) are bounded, they obey a large deviation theorem, the Azuma-Hoeffding inequality, which we use to derive Eq.~(\ref{7)}). Martingales were introduced in quantum information in the context of Bell tests in order to show that having an internal memory of past events is only of marginal help to local models~\cite{ap}{apgill1,apgill2}. Here we use the theory of martingales in the context of randomness generation to show that the same device can be used $n$ times in succession without significantly changing the rate of randomness generation with respect to the mathematically simpler but impractical case where $n$ independent devices are used in parallel.}~\cite{ap}{apgrimmett} with respect to the sequence $\{ W^{k} \, :\, k\ge 2\} $.

The range of the martingales increment are bounded by $|I_{i} -I(W^{i} )|\le \frac{1}{q} +I_{q} $, where $I_{q} $ is the highest possible violation of the inequality $I$ allowed by quantum theory.  We can therefore apply the Azuma-Hoeffding inequality~\cite{ap}{apgrimmett,aphoeffding,apazuma,apalon} $P(Z^{n} \ge \epsilon' )\le \exp \left(-\frac{{\epsilon'}^{2} }{2n(1/q+I_{q} )^{2} } \right)$,  which implies that (with $\epsilon'=n\epsilon$)
\begin{equation} \label{7)}
P\left(\frac{1}{n} \sum _{i=1}^{n}I (W^{i} )\le \frac{1}{n} \sum _{i=1}^{n}I_{i}  -\epsilon\right)\le \delta ,
\end{equation}
where
\begin{equation} \label{8)}
\delta =\exp \left(-\frac{n{\epsilon}^{2} }{2(1/q+I_{q} )^{2} } \right).
\end{equation}
Thus the sum of the Bell expressions $\frac{1}{n} \sum _{i=1}^{n}I (W^{i} )$ can be lower than the observed value $\hat{I}=\frac{1}{n} \sum _{i=1}^{n}I_{i}  $ up to some $\epsilon$ only with some small probability $\delta $. Combining this last result with Eq. \eqref{eq5}, we conclude that
\begin{equation} \label{9)}
H_\infty(R|S)\ge -\log _{2} P(a^{n} b^{n} |x^{n} y^{n} )\ge nf\left(\hat{I}-\epsilon\right)
\end{equation}
with probability at least $1-\delta $.

\subsection{Bounds using no-signalling only}\label{nosig}
As mentioned in the main text, it is also possible to prove lower bounds on the min-entropy of the form Eq.~\eqref{eq1}, but where we impose rather than the full quantum formalism only the no-signalling conditions. That is, we impose that the measurement outcomes produced by the devices cannot be used for arbitrary fast communication, even at a hidden level. We are thus led to a new optimisation problem, which replaces the optimisation problem Eq. \eqref{eq2} we studied previously:
\begin{equation} \label{eq12}
\begin{array}{rcc} {P^{*} (ab|xy)=} & {\max } & {P(ab|xy)} \\ {} & {\text{ subject to}} & {\sum _{abxy}c_{abxy}  P(ab|xy)=I} \\ {} & {} & {\begin{array}{c} {P(ab|xy)\ge 0,} \\ {\sum _{ab}P(ab|xy) =1,} \\ {\sum _{a}P(ab|xy) =P(a|x),} \\ {\sum _{b}P(ab|xy) =P(b|y),} \end{array}} \end{array}
\end{equation}
where the last two equalities are the no-signalling conditions. Solving (\ref{eq12}) (which can be done using linear programming), one deduces a bound of the form $P^{*} (ab|xy)\le \alpha I+\beta $, where in full generality $\alpha $ and $\beta $ may depend on $a,b,x,y$. This follows from the fact that the constraints in eq. \eqref{eq12} define a polytope.

In analogy to Eq.~\eqref{eq3}, one could also consider the maximum of $P(a|x)$ subject to the conditions enumerated in Eq. \eqref{eq12}, which we denote $P^{*} (a|x)$. In this case also the maximum is bounded by a finite set of linear inequalities.

In the case of the CHSH expression, the relevant inequalities are:
\begin{equation} \label{GrindEQ__13_}
P^{*} (a|x)=P^{*} (ab|xy)\le \frac{3}{2} -\frac{I}{4}
\end{equation}
This follows easily by recalling that the extremal points of the no-signalling polytope when $a,b,x,y\in \{ 0,1\} $ fall into two classes, the deterministic points for which $P(ab|xy)\le 1$ and $I=\pm 2$, and the Popescu-Rohrlich boxes for which $P(ab|xy)\le 1/2$ and $I=\pm 4$. This in turn immediately implies a bound of the form eq. \eqref{eq3}:
\begin{equation} \label{GrindEQ__13b_}
H_\infty(AB|XY)\ge -\log _{2} \left(\frac{3}{2} -\frac{I}{4} \right)\,.
\end{equation}
Using this bound, one can now apply directly all the results presented in section \ref{n} to obtain a bound on the min-entropy when the devices are used $n$ times in succession. The only modification is that in Eq. (\ref{8)}), the maximal possible violation $I_q$ of the Bell inequality allowed by quantum theory, must be replaced by the maximal possible violation $I_{ns}$ allowed by no-signalling.

\section{Quantum randomness expanders}\label{diqre}
We now discuss how the results that we have just derived can be used in the context of randomness expansion. Relations between randomness and Bell inequality violations have been discussed in
different contexts in several works, see for instance~\cite{ap}{apvalentini,apconway,apbarrett}. In his PhD thesis~\cite{ap}{apcolbeck}, Colbeck  proposed to exploit this connection for
the task of private randomness generation and formalized this idea concretely. Moreover, he introduced a protocol for
randomness expansion based on Greenberger-Horne-Zeilinger (GHZ)
multipartite pure-state correlations. Following \cite{ap}{apcolbeck}, we define quantum private randomness expansion (QPRE) as a single-party task where a user, Alice, expands a small random initial seed\footnote{Any device-independent random number generation protocol must necessarily use an initial random seed. Indeed, if Alice follows a deterministic procedure, then there exists a predetermined set of measurement outcomes that will pass all the tests performed by Alice, and the adversary can simply provide devices that generate these classical outputs.  Thus the fact that we need an initial random seed is not a weakness of our scheme but a necessity inherent to the concept of device-independent RNG itself.} into a larger string which is close to uniformly random, even when conditioned on the information of a potential quantum adversary. In detail QPRE is realised as follows:
\begin{enumerate}
\item  The setting: Alice has a pair of devices, each of which has $m$ inputs and can produce $d$ outputs. We assume that Alice starts the protocol with an initial private random seed $t=(t_{1},t_{2} )$ divided in two strings $t_{1}$ and $t_{2} $. At the beginning of the protocol, Alice also chooses a security parameter $\delta $ that bounds the probability with which an adversary can cheat.

\item  Use of the devices: Alice uses the string $t_{1} $ to generate pairs of inputs $s=(x_{1} ,y_{1};\ldots;x_{n} ,y_{n} )$. Every pair is generated independently with a probability distribution $P(x,y)$.  Alice then introduces the inputs $(x_{i} ,y_{i} )$ in her devices and obtains the outputs $(a_{i} ,b_{i} )$; she repeat this last step $n$ times for $i=1,\ldots ,n$.

\item  Evaluation of the min-entropy: from the final input and output strings $s=(x_{1} ,y_{1};\ldots;x_{n} ,y_{n})$ and $r=(a_1,b_1;\ldots a_n,b_n)$ Alice computes $\hat I$ (see \eqref{eq6}), and from the chosen value for the security parameter $\delta $, Alice computes $\epsilon$ (see \eqref{8)}). From these values Alice then computes the bound \eqref{9)} on the min-entropy $H_\infty(R|SE)$ of the raw output string $r$.

\item  Randomness extraction: Alice uses a randomness extractor\footnote{Randomness extractors are classical functions that take as input a small uniform random string $t_2$ and a much longer string whose min-entropy is lower bounded by $H_\infty$, and produce as output a string of length $H_\infty$ which is nearly uniform.} and the string $t_{2} $ to convert the raw string $r$ into a slightly smaller string $\bar{r}$ of size $O(nf(\hat{I}-\epsilon))$ which is close to uniform and uncorrelated to the adversary's information.
\end{enumerate}

The final random string $(t,\bar{r})$ of Alice is clearly longer than the initial string $t$ provided that the bound on the entropy $H_\infty(R|SE)$ is strictly positive. However, when the number $n$ of uses of the devices is sufficiently large, it is possible to start from an initial seed of length $O(\sqrt{n} \log _{2} \sqrt{n} )$ to produce a much longer final string of length $O\left(n\right)$. Indeed, if Alice selects one of the $d^{2} $ possible input pairs ($x,y)$ with probability $1-(d^2-1)q$ and the remaining $d^{2} -1$ ones with probability $q$, with $q$ small, the randomness required to generate the inputs is then equal to $nO(-q\log _{2} q)$. Taking $q=O(1/\sqrt{n} )$, the initial random seed $t_{1} $ must thus be of length $O(\sqrt{n} \log _{2} \sqrt{n} )$.  The randomness of the raw string $r$, on the other hand, is given by \eqref{9)} where $\epsilon$ is of order $O(1)$. For fixed Bell violation $\hat{I}$, the output entropy is thus linear in $n$. The raw string $r$ can then be transformed into the fully secure random string $\bar{r}$ using a randomness extractor with the help of a small random seed $t_{2} $, say of length $O(\text{poly}\log n)$~\cite{ap}{apnisan}. This shows that it is possible to construct device-independent QPRE producing $O\left(n\right)$ random bits from an initial seed of total length $O(\sqrt{n} \log _{2} \sqrt{n} )$.

The security of this protocol only holds for the moment against quantum adversaries that measure their quantum side-information before the randomness extraction step. That our analysis holds in this situation follows from the fact that when the adversary performs a measurement on his system, the outcome that he obtains amounts to the preparation of Alice's devices in some particular state. But the bound \eqref{9)} is independent of the way in which the devices have been prepared and thus also holds when conditioned on the outcome of the adversary's measurement.

Our protocol is thus not yet proven to be universally-composable against a full quantum adversary, that is, secure against an adversary that stores his side-information in a quantum memory which can be measured at a later stage. A universally-composable proof would cover the situation in which the adversary tries to estimate the random numbers after getting partial information about them. To obtain a full universally-composable proof of security of our QPRE protocol, a bound on the entropy of the raw string conditioned on the quantum information of an adversary must be derived. Such a bound could then be used in a randomness extraction procedure secure against a fully quantum adversary \cite{ap}{apvidick}.

Proving universally-composable security of our protocol might also open the possibility of more elaborated protocols where Bell violating devices are used in a concatenated way: the random string produced by a first device is used as seed for a second device, whose input is in turn used as seed for the first and so on. Such more complex solutions could lead to much more efficient QPRE devices. Related to this efficiency question, we note that the results of \cite{ap}{apmasanes} and \cite{ap}{aphanggi} imply that exponential randomness expansion, i.e., the generation of an output random string of size $O(n)$ starting from a seed of size $O(\text{poly}\log n)$, is possible in the situation where the devices have no classical or quantum memory. Note, however, that the fact that our bound~(3) holds for devices that have an internal memory is the crucial feature that makes our protocol practical. It would be interesting to show that exponential randomness expansion is also possible in this situation.

Finally, we mention that although we focused here our theoretical analysis on a bipartite protocol (which is easier to implement in practice), our
theoretical approach can also probably be generalized to compute a bound on the randomness produced by the
randomness expansion protocol based on GHZ multipartite correlations introduced in \cite{ap}{apcolbeck}.

\section{Requirements on the devices}\label{loop}
In this section, we discuss the requirements that are necessary to generate random numbers certified by a Bell inequality violation and clarify the concept of device-independence. We also discuss these requirements in the context of the security of the QPRE task and in the context of our experimental demonstration.

We recall that the QPRE system is composed of two devices, referred to in the follows as device $A$ and device $B$. The derivation of the min-entropy bound (3) rests on the following assumptions.
\begin{enumerate}
\item \emph{Quantum-theory: The two devices behave according to quantum theory.}\\ That is when an input $x$ ($y$) is introduced in a device, an output $a$ $(b)$ is produced according to some quantum process.
\item \emph{Input randomness: The inputs $x_i,y_i$ at round $i$ are generated by random processes that are independent and uncorrelated from the devices and their values is revealed to the devices only at round $i$.}\\ This guarantees that the behaviour of the devices at round $i$ does not depend on the values of the inputs used in later rounds, i.e., that the following condition is satisfied
\be \label{cond2}
P(a_ib_i|x_iy_i\ldots x_ny_nW^i)=P(a_ib_i|x_iy_iW^i)\,,
\ee
where $W^i$ denotes, as in Section~A.2, all events in the past of round $i$.  This condition is used in the first line of Eq.~(\ref{eq4}).
\item \emph{Locality: From the moment that the inputs are introduced until the outputs are produced, the two devices are separated and no signal (quantum or classical) can travel from one device to the other. Moreover, the input $x$ is introduced only in device $A$ and the input $y$ only in device $B$.}\\ This assumption ensures that the Hilbert space describing the two devices factorizes as a tensor product and that the measurement operators act on each factor in a way that depends only on the corresponding local input. That is, that the probabilities characterizing a single-run of the experiment are of the form
\be\label{prod}
P(ab|xy)={\text{tr}}\left(\rho M_{x}^{a} \otimes M_{y}^{b} \right)\,.
\ee
This mathematical condition is used in the optimization problem (\ref{eq2}) in Section~A.1.
\end{enumerate}
When the above three conditions are satisfied, the min-entropy bound~(3) derived in Section A follows without additional assumptions on the quantum states, the measurements, the dimension of the Hilbert space, or any other aspects concerning the internal working of the quantum devices. This is what we refer to as \emph{device-independence} --- the fact that the min-entropy bound or the security proof does not depend on a model of the devices.

The conditions listed above represent minimal requirements to apply the device-independence proof in practice: if one of them is not satisfied then it is no longer possible to guarantee the presence of randomness through the violation of a Bell inequality~\footnote{Note that although these three conditions are necessary and that we cannot go completely without them, they can probably be weakened at the expense of the randomness generation rate. This is clear for the first assumption, as we have seen in Section~A.3 that the validity of quantum theory can be replaced by the no-signalling principle. But it might also be possible to weaken the two other assumptions and show that Bell-violating devices can generate randomness even if they have a limited amount of prior information about the future inputs or if they can exchange some limited communication.}. Whether these conditions are verified in an implementation is therefore an important question. However, it is not a question that has a unique and well-defined answer. Depending on the application, the adversarial scenario under consideration, and our level of ``trust" in the devices, there may be different ways to enforce these conditions or verify that they are satisfied. As an illustration, we discuss below three possible ways to ensure that condition~3 is satisfied: i) by imposing space-like separation between the measurement devices; ii) by shielding the devices; iii) by proper design of the setup.

i) Strict space-like separation. Condition 3 states that the devices should be separated and non-interacting during the measurement process. Obviously strict space-like separation between the devices could be used to enforce this separability condition. This space-like separation is required to close the locality loophole in standard Bell tests, where the objective is to disprove alternative theoretical models of Nature that can overcome the laws of physics as they are currently known. Here, however, we assume from the beginning the validity of quantum theory and we use Bell inequalities not as a test of hidden variable models but as a tool to quantify the randomness of quantum theory. Once we assume quantum theory, they are many ways to ensure that the two systems are not interacting other than placing them in space-like intervals, e.g. by shielding the devices as discussed below in point ii).

The view that there is no satisfying way to prevent information from being exchanged between two systems except by strict space-like separation is very conservative. The main problem with this level of paranoia is that it makes cryptography impossible. Indeed, a basic assumption behind all cryptography (classical or quantum) is that users must be confident that information cannot leak out of their location, devices, computers, etc. If the only constraints on the flow of information are those which follows from special relativity, cryptography becomes impossible: any secret bit or key inevitably leaks out to the adversary at the speed of light as soon as produced.

The space-like separation approach may be relevant for non-cryptographic applications of random numbers, where privacy is not an issue, such as Monte Carlo algorithms. Its main advantage is that the condition (\ref{prod}) follows immediately. A secondary point to take into account in this approach, however, is that the inputs $x$ and $y$ must be generated locally and immediately before being introduced into the devices, so that they are prevented from knowing the other device's input\footnote{This implies that the inputs must be generated with a product probability distribution $P(x,y)=P(x)P(y)$. In particular, the joint input distribution cannot be of the form $P(00) = 1-3q, P(01) = P(10) = P(11) = q$ which leads to the quadratic catalysis effect discussed in the previous Section and illustrated in Figure 3. A weaker catalysis effect is nonetheless possible if each individual input is chosen with probabilities $P(0)=1-q, P(1)=q$, resulting in the joint probability $P(00)=(1-q)^2,P(01)=P(10)=q(1-q),P(11)=q^2$. Taking $q^2=O(n^{1/4})$, we find that $\epsilon$ in Eq.~(\ref{9)}) is of order $O(1)$, and thus that the output randomness is of order $O(n)$. The randomness consumed at the inputs, on the other hand, is $nO(-q\log_2q)=O(n^{3/4}\log_2n^{1/4})$.}.

Note that space-like separation is essential to test more fundamental implications of our results, such as the compatibility between determinism and no-signalling mentioned at the end of the paper.

ii) Shielding of the devices. Here, condition 3 is enforced by separating and shielding the devices and the relevant part of the laboratory. This view is consistent with the known forces in physics which are either extremely weak (gravity), short range (weak and strong force), or can be screened by conductors (electro-magnetic force). Note that the shields must have doors with allow the boxes to interact with the external world (dotted lines in Figure~1). The doors must allow both classical and quantum information through, to let the devices  establish shared entanglement, receive the inputs $x,y$, and produce the outputs $a,b$.

The assumption that laboratories and devices can be adequately shielded against unwanted leakage of information is implicit in all of cryptography. From the basic cryptographic requirement that the two devices can be adequately shielded so that they do not send information to the adversary, it is only a small extra step to require that they can also be adequately shielded so that they cannot signal to each other. It would be very artificial to assume that the devices are not able to communicate to the adversary, an essential requirement for any cryptographic application, but can communicate between each other.

The main disadvantage of this approach is that we are making a technological assumption. One can never completely rule out that the shielding is sufficient. Particular attention would have to be paid to this point if it is an adversary that has provided the devices.

iii) Proper design of the experiment. We may have a good confidence that condition~3 is satisfied simply based on a proper design and superficial description of the devices. For instance in the experiment reported here we are using two atoms that are confined in two independent vacuum chambers separated by about 1 meter. At this distance, direct interaction between the atoms is negligible and classical microwave and optical fields used to perform measurements on one atom have no influence on the other atom. Based on this superficial description of the setup, we can safely assume that we are dealing with two independent quantum systems and that the general formalism used to derive our bound applies\footnote{At first sight a better way to implement our proposal experimentally would be to use two ions in the same trap, as in \cite{ap}{aprowe}, as data rates orders of magnitude higher can be achieved. But in this case we cannot assume that the two ions are separate and non interacting without modeling the details of the experiment. Indeed the ions are at all times strongly coupled through the vibrational modes of the cavity, and furthermore the measurements cannot be assumed to act on each ion separately as they are separated by a few microns, comparable to the wavelength of light used for readout.}.

Obviously, this approach is insecure in the adversarial scenario where the devices are untrusted and assumed to be prepared by an adversary, in which case it is necessary to take active measures to control the flow of information in and out of the devices, as in point ii) above. The application of the concept of device-independence, however, is not restricted to this strong adversarial scenario with untrusted devices. It may be perfectly legitimate to assume that the devices have been built in good faith and that they do not contain any sneaky component sending unwanted signals (we must anyway trust the classical computers, devices, etc., used to analyze the data). The problem, however, is that it is very difficult, even for honest parties, to construct reliable random number generators. The theoretical modeling of the physical process at the origin of the randomness source is often incomplete and relies on assumptions which may be very difficult to confirm.  This makes an accurate estimation of the entropy produced by the devices difficult. Most RNGs break silently, often producing decreasingly random numbers as they degrade. Failure modes in such devices are plentiful and are complicated, slow, and hard to detect. Imperfections can introduce patterns undetected by statistical tests but known to an adversary. The generation of randomness in a device-independent way (that is without detailed modeling of the devices) solves all these shortcomings of traditional random number generators: the violation of a Bell inequality provides a bound on the randomness produced by the devices, independently of any noise, imperfections, lack of knowledge, or limited control of the apparatuses.

Considerations similar to the above ones can be made for the second condition in our list of assumptions, i.e., the requirement that the inputs should be chosen in a way that appears ``random" to the devices. In our experiment, the value of the chosen measurement bases $(x_i,y_i)$, obtained by combining the outputs of several (public) random number generators, is unlikely to be correlated to the state of the atoms before the measurement microwave pulses are applied, and we can safely assume that Eq.~(\ref{cond2}) is satisfied.  Of course in a strong adversarial scenario where the devices are untrusted, the adversary could exploit a prior knowledge of the inputs in the design of the devices. In this case, it is thus important to generate the inputs with a private random source.

Finally, it is tacit, but obvious, that the implementation should conform to the theoretical description detailed in Section~B. For instance, in the case of the CHSH inequality, we consider devices that produce binary outcomes $a,b\in\{0,1\}$ and use all the raw data produced over the $n$ runs of the experiment in our analysis. In photonic Bell tests, detectors have low efficiencies, and device often produce a no-event outcome $\bot$ corresponding to the absence of a click, i.e., $a,b\in\{0,1,\bot\}$. This leads to the famous detection loophole which is circumvented by applying a post-selection to the observed data. This represents a departure from our description and our results can therefore no longer be applied in this case. In our experiment, this problem is not an issue, as every event is recorded.

More generally, it is straightforward that a Bell violation without closing the detection loophole (or other loopholes arising from post selection, such as the coincidence time loophole \cite{ap}{larsson}) cannot be used to certify randomness generation. In a scenario where the devices are trusted, one could address this requirement by means of the fair sampling assumption, which states that the detected events represent a fair sample of all the events (including the undetected ones). However, making the fair sampling assumption requires a detailed knowledge of the devices, and could fail without one being aware of it (see also attacks based on the detection loophole in traditional QKD \cite{ap}{lo1,lo2}.).

\section{Experiment}
\subsection{Experimental system}\label{exp}
Individual $^{171}$Yb$^+$ atomic ions are stored in two rf Paul traps located in two independent vacuum chambers separated by about 1 meter and placed in the magnetic field of 3.4 G that defines the quantization axis. We encode qubits in the $F=1$, $m_F=0$ and $F=0$, $m_F =0$ hyperfine levels of the $S_{1/2}$ ground state of each atom, denoted by \textbar 1$\rangle$ and \textbar 0$\rangle$, respectively. Each experimental sequence begins with a 1 $\mu$s optical pumping pulse and 10 $\mu$s microwave pulse that prepares each atom in the \textbar 0$\rangle$+\textbar 1$\rangle$ state. Next, a fast laser pulse linearly polarized along the direction of magnetic field excites both ions to the $P_{1/2}$ state with a probability close to unity.  If the atoms each return to the ground state by emission of a $\pi $-polarized photon, the frequency of the emitted photons are entangled with the atomic qubit states.  With a small probability, both photons are collected from each atom and interfered on a 50:50 beamsplitter, with polarizing filters selecting only $\pi$-polarized light. Detection of coincident photons behind the beamsplitter heralds the preparation of the entangled state of the two atoms: $\left|\left.0\right\rangle \right|\left.1\right\rangle -e^{i\chi }|\left.1\right\rangle |\left.0\right\rangle $~\cite{ap}{apmaunz,apolmschenk}. The experiment is repeated at 95 kHz, but due to finite collection and detection efficiency of the photons, we observe on average only 1 entanglement event every 8 minutes.  Given a successful entanglement event, we finally detect each qubit in measurement bases determined by the input variables  $x$ and $y$~\cite{ap}{apolmschenk2}. Before measurement, microwave pulses with a controlled phase and duration resonant with the $|0\rangle\rightarrow|1\rangle$ transition rotate each atomic qubit through a polar angle of $\pi/2$ on the Bloch sphere.  The phase difference $\varphi_x-\varphi_y$ between the rotations on the two atoms is adjusted with respect to the phase $\chi $ of the entangled state so that the qubits are appropriately rotated for a Bell inequality violation~\cite{ap}{apmatsukevich}.  Finally, each atom is illuminated with laser light resonant with the $S_{1/2}$ $F=1$ $\rightarrow$ $P_{1/2}$ $F=0$ transition, and the \textbar 1$\rangle$ state fluoresces strongly while the \textbar 0$\rangle$ state is nearly dark, resulting in a detection of each qubit state with 98.5\% accuracy \cite{ap}{apolmschenk2}.

\subsection{Generation of measurement settings}\label{meas}
The measurement settings were chosen by combining several online random number generators that use: radioactive decay as a source of randomness~\cite{ap}{aprand1}; atmospheric noise as a source of randomness~\cite{ap}{aprand2}; and randomness derived from remote computer and network activity~\cite{ap}{aprand3}. We requested 128 bytes (corresponding to 32 integer numbers) from each of these generators and combined them using XOR function. This procedure should produce measurement settings that are independent and uniform. Indeed the measurement settings were tested to ascertain that they did not contain any obvious bias (see Appendix \ref{tests}).

Before every entanglement event the two least significant bits of one of the above integer numbers was transmitted to the FPGA board that controls our experimental time sequence. After a two photon coincidence event the board applies microwave pulses to both ions with a phase that depends on the measurement setting it obtains from the computer, counts the number of photons during ions state detection, transmits the state detection results and the measurement settings it used back to the computer, and receives new measurement settings for the next entanglement event. When the number of available random numbers in the buffer approaches 0,  a new block of random numbers is requested from the random number generators and added to the buffer.

\subsection{Comparison with local causality}\label{viol}
The usual interpretation of the experimental violation of a Bell inequality is that the observed correlations cannot be reproduced by a local model~\cite{ap}{apbell}. This should be properly expressed as a statistical statement. Here we follow the approach of \cite{ap}{apgill1,apgill2} (see also Section A.2) to derive the relevant statement.

We consider the quantity $I_{i} $ defined in Eq. \eqref{eq6}, and focus on the CHSH inequality with uniform choice of settings ($P(xy)=1/4$). We denote by the superscript $l$ the predictions of local theories.

Let us consider the predictions of local theories. In this case we have $-2\le E(I^l_i\left(W^i\right)=I^l\left(W^i\right))\le 2$. The random variables${\ Z}^k=\sum^k_{i=1}{\left(\hat I_i-I^l\left(W^i\right)\right)}$ are a martingale with respect to the sequence $\{ W^{k} \, :\, k\ge 2\} $. The range of the martingales increments are bounded by $\left|\hat I_i-I^l\left(W^i\right)\right|\le 4+2=6$, where the 4 comes from the possible range of $\hat I_i$  and the 2 comes from the range of $I^l\left(W^i\right)$. Applying the Azuma-Hoeffding inequality then implies that the probability that a local deterministic theory will yield a violation greater or equal than the observed violation $\hat{I}=\frac{1}{n} \sum _{i=1}^{n}I_{i}$ is bounded by:
\be
P(\frac{1}{n}\sum^n_{i=1}{I^l(W^i)\ge \hat I)\le \text{exp}}\left(-\frac{n{\left(\hat I-2\right)}^2}{72}\right).\ee
This inequality holds for all local models (even if they can exploit an internal memory that remembers past measurement settings and past outputs) which attempt to violate the CHSH inequality (with the measurement settings uniformly distributed).
Inserting the data from our experiment (Table 1) shows that the probability that the measurement results could have been produced by a local model is less than 0.000767.

\subsection{Statistical tests}\label{tests}
\noindent As a check of consistency and of the quality of our experiment, we verified that the bit strings $(x_1,y_1;\ldots;x_n,y_n)$ used as measurement settings and the bit strings $a_1,b_1;\ldots;a_n,b_n$ produced as outputs did not reveal any obvious non-random pattern using a standard battery of statistical tests~\cite{ap}{apmarsaglia,aprukhin,apmenezes}. However, those batteries were originally designed for sequences containing millions of bits. As the strings we consider contain several thousands of bits, only the tests statistically relevant for such small strings were performed. Specifically the tests called ``Frequency'', ``Block frequency'', ``Runs'', ``DFT'', ``Serial'' and ``Approximate Entropy'' were reprogrammed from~\cite{ap}{aprukhin} (sections 2.1, 2.2, 2.3, 2.6, 2.11, 2.12 ) and the tests called ``Two-bit'' and ``Poker'' were reprogrammed from~\cite{ap}{apmenezes} (section 5.4.4.).

The result of these statistical tests are described by a p-value, which is the probability that a perfect random number generator would have produced a sequence less random than the sequence that was tested, given the kind of non-randomness assessed by the test. A p-value of 1 therefore means that the sequence tested appears to have been generated by a perfectly random process, while a p-value of 0 implies that the sequence appear to be completely non-random~\cite{ap}{aprukhin}. Together with computing the p-value, a significance level $\alpha $ should be chosen for the tests. If the p-value $\geq$ $\alpha $, then the hypothesis that the sequence appears to be random is accepted. If the p-value $<$ $\alpha $, then this hypothesis is rejected, i.e., the sequence appears to be non-random. Typical values for $\alpha $ are in the range [0.0001, 0.01] and we choose $\alpha $ =0.001. An $\alpha $ of 0.001 indicates that one would expect one sequence in 1000 sequences to be rejected by the test if the sequence was generated by a perfect random number generator. For a p-value $\geq$ 0.001, a sequence would be considered to be random with a confidence of 99.9\%. For a p-value $<$ 0.001, a sequence would be considered to be non-random with a confidence of 99.9\%~\cite{ap}{aprukhin}.

The results of these statistical tests (given by the p-value) evaluated on our data are given in Table~2. As in Section~A.2, we use the notation $x^{n} =(x_{1} ,...,x_{n} )$, $a^{n} =(a_{1} ,...,a_{n} )$ for the settings and outputs for the first ion; and $y^{n} =(y_{1} ,...,y_{n} )$, $b^{n} =(b_{1} ,...,b_{n} )$ for the settings and outputs for the second ion. Each of these strings contains $n=3016$ bits and the combination $(a_{1} ,b_{1} ,...,a_{n} ,b_{n} )$ contains therefore $6032$ bits.
\setcounter{table}{1}\begin{table}[t]\begin{center}
\begin{tabular}{@{\extracolsep{1em}}|c|c c c c|} \hline
Statistical test & $x^{n} y^{n} $ & $a^{n} $ & $b^{n} $ & $(a_{1} ,b_{1} ,...,a_{n} ,b_{n} )$ \\ \hline
Frequency & 0.9795 &\quad 0.7988 & 0.0743 & 0.1493 \\
Block Frequency & 0.2668 & 0.5540 & 0.0592 & 0.1360 \\
Runs & 0.6616 & 0.2904 & 0.4312 & 0.0000 \\
DFT & 0.8542 & 0.3096 & 0.8686 & 0.3081 \\
Serial & 0.1681 & 0.1300 & 0.0119 & 0.0000 \\
Approximate Entropy & 0.2445 & 0.5568 & 0.0130 & 0.0000 \\
Two Bit & 0.9024 & 0.5699 & 0.1579 & 0.0000 \\
Poker & 0.0189 & 0.1933 & 0.0353 & 0.0000 \\ \hline
\end{tabular}\caption{Results of the statistical tests given by the p-values evaluated on the measurement settings, the outputs of one atom and the other taken separately, and the outputs of the two atoms taken together.}
\end{center}
\end{table}

As we see from the data in Table~2, the input string $x^{n} y^{n} $ passes all the tests since the p-values are larger than the chosen significance level $\alpha $ =0.001. This is expected as the random number generators we used have already been extensively tested. The outputs strings $a^{n} $ and $b^{n} $ taken separately also pass all the tests. This confirms that the experiment performed well and did not have any important internal bias.  Finally, as expected, if all the outputs are grouped into the combination $(a_{1} ,b_{1} ,...,a_{n} ,b_{n} )$, then the results do not pass the tests because the measurement outputs on the two atoms are correlated and thus are not independent random variables.

Note that the general theory described in this paper guarantee that the measurement outputs contain at the 99\% confidence level 42 new random bits, independently of any hypothesis on how the experiment was realised. This is a much stronger statement than passing or not passing the statistical tests mentioned above, which merely indicate that no obvious non-random patterns are present in the measurement outputs.

To verify that the atoms did not influence each other during the measurement in an obvious way, we also performed statistical tests to check that the data reported in Table 1 in the main text are compatible with the no-signalling conditions given in Eq. \eqref{eq12}.  There are 4 no-signalling conditions to check. In each case we performed a two-sided Fisher's test. We found p-values of 0.7610; 0.7933; 0.7933; 0.2867 which show that the data are compatible with the no-signalling hypothesis.

\bibliographystyle{ap}{unsrt}
\bibliography{ap}{randomness_ap}{Appendix references}

\end{document}